\documentclass[11pt]{article}
\usepackage{bm}
\usepackage{graphicx}
\usepackage{amssymb,amsmath}
\usepackage{multirow}
\usepackage{cite,color,url}
\usepackage{subcaption}
\usepackage[colorlinks=true
,urlcolor=blue
,anchorcolor=blue
,citecolor=blue
,filecolor=blue
,linkcolor=blue
,menucolor=blue
,linktocpage=true
,pdfproducer=medialab
,pdfa=true3
]{hyperref}

\usepackage{slashed}
\usepackage{epsfig,psfrag,rotating,soul}
\usepackage{rotfloat}
\usepackage[font={small}]{caption}
\usepackage{colortbl}
\usepackage{xcolor}
\usepackage{comment}
\definecolor{darkspringgreen}{rgb}{0.09, 0.45, 0.27}

\usepackage{extarrows} 

\evensidemargin \oddsidemargin
\allowdisplaybreaks

\let\OLDthebibliography\thebibliography
\renewcommand\thebibliography[1]{
  \OLDthebibliography{#1}
  \setlength{\parskip}{0pt}
  \setlength{\itemsep}{0pt plus 0.3ex}
}

\definecolor{mygray}{gray}{0.85} 
\definecolor{myblue}{cmyk}{0.65, 0.37, 0.0, 0.19}

\begin{document}
\thispagestyle{empty}

\def\thefootnote{\fnsymbol{footnote}}

\vspace*{1cm}

\begin{center}

\begin{Large}
\textbf{\textsc{{Rescuing Overabundant Dark Matter with a  Strongly First Order Phase Transition in the Dark Sector}}}
\end{Large}


\vspace{1cm}

{\sc
 Peisi~Huang$^{1, 7}$%
\footnote{{\tt \href{mailto:gabriel.zapata@pucp.edu.pe}{peisi.huang@unl.edu}}}%
, Anibal~D.~Medina$^{2}$%
\footnote{{\tt \href{mailto:anibal.medina@fisica.unlp.edu.ar}{anibal.medina@fisica.unlp.edu.ar}}}%
 and Carlos~E.~M.~Wagner$^{3, 4, 5, 6, 7}$%
\footnote{{\tt \href{mailto:cwagner@anl.gov}{cwagner@uchicago.edu}}}%

}

\vspace*{.7cm}

{\sl
$^1$Department of Physics and Astronomy, University of Nebraska, Lincoln, NE 68588, USA

$^2$IFLP, CONICET - Dpto. de F\'{\i}sica, Universidad Nacional de La Plata,  C.C. 67, 1900 La Plata,  Argentina

\vspace*{0.1cm}

$^3$HEP Division, Argonne National Laboratory, 9700 Cass Ave., Argonne, IL 60439, USA

\vspace*{0.1cm}

$^4$Enrico Fermi Institute, Physics Department, University of Chicago, Chicago, IL 60637, USA

\vspace*{0.1cm}

$^5$Kavli Institute for Cosmological Physics, University of Chicago, Chicago, IL 60637, USA

\vspace*{0.1cm} 

$^6$ Leinweber Institute for Theoretical Physics, University of Chicago, Chicago, IL 60637, USA

\vspace*{0.1cm}

$^7$ Perimeter Institute for Theoretical Physics, Waterloo, 
ON N2L 2Y5, Canada

}

\end{center}

\vspace{0.1cm}

\begin{abstract}
\noindent

We consider a dark sector consisting of fermionic dark matter~(DM) charged under a broken dark $U(1)_D$ gauge symmetry, interacting with the Standard Model through kinetic mixing. In such models, the DM annihilation cross section is typically suppressed by the small kinetic mixing and or a heavy mediator, often leading to an overabundant relic density. We show that the observed DM abundance can be achieved if the dark Higgs undergoes a strong first order phase transition after DM freeze-out. In this scenario, the relic abundance is set by thermal freeze-out in the symmetric phase and subsequently reduced by entropy injection from the phase transition, rather than by annihilation in the broken phase. We find that to reproduce the observed relic abundance, the required phase transition is generically supercooled. The resulting stochastic gravitational wave signal lies within the sensitivity of future experiments, providing a complementary probe of this framework. Moreover, a strongly supercooled phase transition can potentially account for the NANOGrav signal for DM masses below $O(10)$ GeV.

\end{abstract}

\def\thefootnote{\arabic{footnote}}
\setcounter{page}{0}
\setcounter{footnote}{0}

\newpage

\section{Introduction}
\label{intro}

The existence of dark matter~(DM) is firmly established through a wide range of astrophysical and cosmological observations~\cite{Zwicky,Rubin:1970zza,1980ApJ...238..471R,1978PhDT.......195B}, yet its particle nature remains unknown. This has motivated an extensive experimental and observational program, including direct detection, indirect detection, collider searches, and cosmological probes, aimed at probing the particle nature of DM~\cite{Bertone:2004pz,Schumann:2019eaa,Gaskins:2016cha,Boveia:2018yeb}. A simple and well motivated class of scenarios introduces a dark sector charged under an additional Abelian gauge symmetry, with interactions between the dark sector and the Standard Model~(SM) mediated by kinetic mixing between the dark photon and hypercharge. Such dark photon models arise naturally in many extensions of the SM and provide a framework in which DM interactions with visible sector that is consistent with current experimental constraints~\cite{Pospelov:2007mp, Ackerman:2008kmp, Chu:2011be,Essig:2013lka}.

In the standard thermal freeze-out picture, obtaining the observed DM abundance requires a velocity-averaged annihilation cross section of $\langle \sigma v \rangle \simeq 3\times 10^{-26}\, \textrm{cm}^3\textrm{s}^{-1}$. If the annihilation cross section is significantly smaller than this value throughout the thermal history, DM would be overproduced in the early Universe.

In this work, we consider a dark sector consisting of a dark Higgs field, a dark photon kinetically mixed with the SM hypercharge, and Dirac fermionic DM~\cite{Fabbrichesi:2020wbt,Curtin:2014cca, Alenezi:2025kwl}. We focus on regions of parameter space in which the DM annihilation cross section at late times is strongly suppressed and would therefore be naively interpreted as  an overabundant relic density. In dark photon models, such suppression typically arises from a heavy mediator and/or a small kinetic mixing parameter, as is often required by direct detection constraints for DM masses above $O(10)$ GeV. For lighter DM, the annihilation cross section at late times is instead constrained by observations of the cosmic microwave background~(CMB), which impose strong limits on energy injection after recombination and require $\langle \sigma v\rangle$ today to be well below the thermal target~\cite{PhysRevD.87.123513,Slatyer:2015jla}. In both cases, these considerations point to the same naive conclusion, namely that a suppressed annihilation rate is incompatible with the observed DM abundance with a standard thermal history.


Instead, we discuss the possibility that the observed DM relic abundance is obtained if the dark Higgs field undergoes a strong first order phase transition in the early Universe. Before the phase transition, the dark $U(1)_D$ symmetry is unbroken and the dark photon is massless, so DM annihilation is dominated by the unsuppressed t- and u-channel processes into pairs of dark photons. The freeze-out therefore occurs in this symmetric phase, efficiently depleting the abundance. When the phase transition takes place, the dark Higgs acquires a nonzero vacuum expectation value~(vev) and the dark photon becomes massive. If the dark photon mass is larger than the DM mass, annihilation into dark photon pairs is no longer kinematically allowed, and DM annihilation is instead dominated by s-channel processes into SM particles. At late times, this annihilation channel is strongly suppressed by the small kinetic mixing and/or the heavy mediator mass, which is often required by direct detection and cosmological constraints. In the parameter region of interest, the phase transition is generically supercooled. The associated release of vacuum energy and subsequent entropy injection dilute the DM number density. As a result, the present day relic abundance is not determined by the annihilation at the broken phase, but by the combination of that freeze-out earlier in the symmetric phase and the subsequent entropy dilution induced by the supercooled phase transition.

In addition, the first order phase transition generically produces a stochastic gravitational-wave~(GW) background, providing a complimentary observational probe to this scenario. For other scenarios in which GW are produced via phase transitions in the dark sector, see for example~\cite{Allahverdi:2024ofe,Mahapatra:2026fyv,Adhikary:2024btd,Oikonomou:2024geq, Feng:2026iqo}.

This paper is organized as follows. In Sec.~\ref{sec:model}, we introduce the dark sector model and its interactions with the SM. In Sec.~\ref{sec:relic abundance}, we analyze the DM relic abundance and show how a late dark Higgs phase transition modifies the thermal history. We discuss experimental and cosmological constraints and define representative benchmark scenarios in Sec.~\ref{sec:constraints and signals}. We then study the GW signal associated with the dark Higgs phase transition, including both heavy and light DM regimes. We conclude in Sec.~\ref{sec:conclusion}.

\section{The model}
\label{sec:model}
We consider a gauge Abelian extension of the SM in which the associated dark photon mixes kinetically with the SM hypercharge gauge boson and couples to a SM-neutral Dirac fermion, associated with the DM. The Abelian gauge symmetry spontaneously breaks (SSB) at some point in the early Universe leading the dark photon to acquire a mass via a phase transition in the dark sector. In the symmetric phase, the relevant part of the zero-temperature tree-level Lagrangian reads as,
\begin{equation}
\Delta\mathcal{L}=-\frac{1}{4}F'_{\mu\nu}F^{\prime\mu\nu}-\frac{\epsilon}{2 c_W}F'_{\mu\nu}B^{\mu\nu}+\bar{\psi}(iD_{\mu}\gamma^{\mu}-m_{\psi})\psi+D_{\mu}\Phi^{\star}D_{\mu}\Phi-V(\Phi) \end{equation}
where $B_{\mu}$ is the SM hypercharge gauge boson and $A'_{\mu}$ denotes the massless dark photon gauge field, $D_{\mu}\equiv \partial_{\mu}-ig_{D}Q_{D}A'_{\mu}$ is the covariant derivative in the dark sector with $g_D$ and $Q_D$ the dark gauge coupling and the charge of the field, $\psi$ is the Dirac DM fermionic candidate with a $\Phi$-independent mass $m_{\psi}$, $\epsilon$ is a dimensionless constant that parametrizes the kinetic mixing and $c_W\equiv \cos\theta_W=g/\sqrt{g^2+g'^2}$, with $\theta_W$ the weak mixing angle, $g$ and $g'$ the $SU(2)_L$ and $U(1)_Y$ gauge couplings, respectively. We include for completion the dark Higgs sector kinetic term and potential, which we leave generic at the moment, though leading to SSB and providing a mass $m^2_{A'}=g^2_D Q^2_{\Phi} v^2_{\Phi}$ to the dark photon. We assume for simplicity no mixing between the dark Higgs and the SM Higgs, and take $Q_{\Phi}=Q_{\psi}=1$. In the case that the electroweak symmetry is also unbroken, an additional orthogonal rotation remains related to the inability of distinguishing between the two massless gauge bosons, rendering the mixing angle a free parameter~\cite{delAguila:1995rb}. Note that the interactions of the neutrally electromagnetic gauge bosons with the fermion currents are $g_D A'_{\mu} J^{\mu}_{D}$, $g Z_{\mu} J^{\mu}_Z$ and $e A_{\mu}J^{\mu}_{EM}$, with $Z_{\mu}=c_W W^3_{\mu}-s_W B_{\mu}$ and $A_{\mu}=s_W W^3_{\mu}+c_W B_{\mu}$, and $J^{\mu}_{D}$, $J^{\mu}_{Z}$ and $J^{\mu}_{EM}$, the dark, weak and electromagnetic currents, respectively.

One can easily remove the kinetic mixing term by a proper shift in either $A'_{\mu}$ or $B_{\mu}$~\cite{Holdom:1985ag,Babu:1997st}. As mentioned before, in the symmetric phase for the dark sector, the mixing angle remains a free parameter. Thus in the following we consider that SSB has already taken place.  In particular, shifting $B_{\mu}=\hat{B}_{\mu}+\gamma A'_{\mu}$, and canonically normalizing $\hat{A}'_{\mu}=A'_{\mu}/\sqrt{1-\gamma^2}$ with $\gamma=\epsilon/c_W$, one can write a canonically well normalized kinetic term for the dark photon and hypercharge as,
\begin{equation}
\Delta\mathcal{L}\supset -\frac{1}{4}\hat{F}'_{\mu\nu}\hat{F}^{\prime\mu\nu}-\frac{1}{4}\hat{B}_{\mu\nu}\hat{B}^{\mu\nu}+D_{\mu}H^{\dagger}D^{\mu}H+\frac{1}{2}m^2_{A'}(1-\gamma^2)\hat{A}'_{\mu}\hat{A}'^{\mu}
\end{equation}
where here $D_{\mu}=\partial_{\mu}-igW^{a}_{\mu}T^{a}-ig'YB_{\mu}$ is the $SU(2)_L\times U(1)_Y$ covariant derivative. We see then that via the shift in the hypercharge field $B_{\mu}$, we've introduced now a mass mixing between the well-normalized fields $\hat{A}'_{\mu}$ and $\hat{B}_{\mu}$, once the dark and Electroweak SM symmetries are spontaneously broken. Assuming that SSB in the dark sector takes place at a lower scale than the Electroweak symmetry breaking scale ($v=246$ GeV), as it will turn out for our supercooled phase transition, the neutral EW gauge bosons $W^{3}_{\mu}$, $B_{\mu}$ will mix with the dark photon. The mass mixing matrix in the basis $(W^{3}_{\mu},\hat{B}_{\mu},\hat{A}'_{\mu})$ takes the form,
\begin{eqnarray}
M= \left(
\begin{array}{ccc}
\frac{g^2v^2}{4} & -\frac{gg'v^2}{4} & -\frac{gg'v^2}{4}\gamma\sqrt{1-\gamma^2}   \\
.  & \frac{g'^2v^2}{4} & \frac{g'^2v^2}{4}\gamma\sqrt{1-\gamma^2}   \\
. & . & m^2_{A'}(1-\gamma^2)+\frac{g'^2v^2}{4}\gamma^2(1-\gamma^2)
\end{array}
\right) . 
\end{eqnarray}
One can easily check that the determinant of this mass mixing matrix vanishes, providing a massless gauge boson associated with the physical photon (QED remains unbroken). We can make this more transparent by rotating the mass matrix $M'=U^{t}.M.U$, with,
\begin{eqnarray}
U = \left(
\begin{array}{ccc}
s_W & c_W & 0   \\
.  & -s_W & 0   \\
. & . &1
\end{array}
\right) . 
\end{eqnarray}
such that the rotated mass matrix takes the form,
\begin{eqnarray}
M' = \left(
\begin{array}{ccc}
0 & 0 & 0   \\
.  & \frac{1}{4}(g^2+g'^2)v^2 & -\frac{g´}{4}\sqrt{g^2+g'^2}v^2 \gamma\sqrt{1-\gamma^2}   \\
. & . & -\frac{1}{4}(1+\gamma^2)(4m^2_{A'}+g'^2v^2\gamma^2)
\end{array}
\right) =  \left(
\begin{array}{ccc}
0 & 0 & 0   \\
.  & m^2_Z & m^2_{ZA'}   \\
. & . & m^2_{\tilde{A}}
\end{array}
\right)
\end{eqnarray}
It is clear that in this rotated basis the eigenstate $A^{mass}_{\mu}=(1,0,0)=s_W W^3_{\mu}+c_W \hat{B}_{\mu}$ is the physical massless photon and  we can focus on diagonalizing the $2\times 2$ submatrix, whose eigenvalues can be written as,
\begin{equation}
m^2_{2,1}=\frac{1}{2}\left(m^2_Z+m^2_{\tilde{A}}\pm\sqrt{(m^2_Z-m^2_{\tilde{A}})^2+4m^4_{ZA}}\right)   
\end{equation}
and whose eigenstates we denote as $\chi_{2\mu}=x_a \tilde{Z}_{\mu}+x_b \hat{A}'_{\mu}$ and $\chi_{1\mu}=y_a \tilde{Z}_{\mu}+y_b \hat{A}'_{\mu}$
, with $\tilde{Z}_{\mu}=c_W W^3_{\mu}-s_W \hat{B}_{\mu}$. It is easy to check that $y_a=-x_b$ and $y_b=x_a$, and that $x_a$ and $x_b$ are given by,
\begin{equation}
    x_a=\frac{m^2_{ZA}}{\sqrt{m^4_{ZA}+(m^2_{2}-m^2_Z)^2}},\;\; x_b=\frac{m^2_{2}-m^2_Z}{\sqrt{m^4_{ZA}+(m^2_{2}-m^2_Z)^2}}
\end{equation}

In order to get the interactions of the gauge mass eigenstates with the SM fermions and the DM, we just need to write the initial states $A'_{\mu}$, $A_{\mu}$ and $Z_{\mu}$ in terms of the gauge mass eigenstates as~\cite{Alenezi:2025kwl}:
\begin{eqnarray}
    A'_{\mu}&=&\sqrt{1-\gamma^2}x_b \chi_{2\mu}+\sqrt{1-\gamma^2}x_a \chi_{1\mu}\\
    Z_{\mu}&=&(x_a-s_W\gamma x_b \sqrt{1-\gamma^2})\chi_{2\mu}-(x_b+s_W x_a\gamma\sqrt{1-\gamma^2})\chi_{1\mu}\\
A_{\mu}&=&A^{mass}_{\mu}+c_W \gamma \sqrt{1-\gamma^2}(x_b \chi_{2\mu}+x_a \chi_{1\mu})
\end{eqnarray}
Recall that these initial gauge states are the ones that couple to the dark, weak and electromagnetic currents, respectively. From these expressions we clearly notice that the physical photon $A^{mass}_{\mu}$ only couples to the electromagnetic current and in particular it does not couple to the DM current.
\section{Relic Abundance with a Late Dark Higgs Phase Transition}
\label{sec:relic abundance}
In this section, we study the thermal history of the DM and determine its relic abundance. At zero temperature, the dark $U(1)_D$ symmetry is spontaneously broken, and both the dark Higgs and the dark photon acquire masses proportional to the vev, $\langle \Phi \rangle=v_{\Phi}/\sqrt{2}$. In particular, the dark photon mass is given by $m_{A'}=g_{D}v_{\Phi}$, while the dark Higgs mass is proportional to the vev, $m_{\varphi}\propto v_{\Phi}$. For a scalar potential with a negative mass squared term and a quartic self-interaction, one has $m_{\varphi}=\sqrt{2\lambda_{\Phi}}v_{\Phi}$, in which $\lambda_{\Phi}$ is the quartic coupling of the dark Higgs. 
In this work, we focus on the region of parameter space in which the DM mass satisfies $m_\psi < m_\varphi$ and $m_\psi < m_{A^{\prime}}$, $m_{A^{\prime}}\neq 2 m_{\psi}$ . In this region, the t-channel annihilation processes $\psi\bar{\psi} \rightarrow A^{\prime} A^{\prime}$ and $\psi \bar{\psi} \rightarrow 
\varphi\varphi$ are kinematically forbidden. As a result, DM annihilates dominantly into SM particles through an off-shell s-channel dark photon, $\psi \bar{\psi} \rightarrow A^{\prime *}\rightarrow f\bar{f}$, in which $f$ denotes the SM fermions. The corresponding annihilation cross section scales as $g_D^2\epsilon^2 m_\psi^2/m_{A^\prime}^4$. In this work, we neglect resonant annihilation through an on-shell dark photon. 
We calculate the thermal average annihilation cross section. We compute the s-channel processes mediated by the neutral state $\chi_{1,2}$ (as defined in Sec~\ref{sec:model}), taking the non-relativistic limit which is appropriate for DM annihilation today. Assuming for simplicity massless SM fermions,
\begin{equation}
\langle \sigma v\rangle= \frac{g^2 g^2_D   m^2_{\psi}}{\pi c^2_W}\sum_{f}\left[\left(\frac{g_{\psi,1} g_{V1,f}}{m^2_{1}-4m^2_{\psi}}+\frac{g_{\psi,2} g_{V2,f}}{m^2_{2}-4m^2_{\psi}}\right)^2+\left(\frac{g_{\psi,1} g_{A1,f}}{m^2_{1}-4m^2_{\psi}}+\frac{g_{\psi,2} g_{A2,f}}{m^2_{2}-4m^2_{\psi}}\right)^2\right]
\end{equation}
where the vectorial, axial and DM couplings take the form,
\begin{eqnarray}
     g_{V1,f}&=&Q_f s_W c^2_W\gamma\sqrt{1-\gamma^2}x_a-(x_b+s_W\gamma\sqrt{1-\gamma^2}x_a)\left(\frac{T_3}{2}-s^2_W Q_f\right)\\
      g_{V2,f}&=&Q_f s_W c^2_W\gamma\sqrt{1-\gamma^2}x_b+(x_a-s_W\gamma\sqrt{1-\gamma^2}x_b)\left(\frac{T_3}{2}-s^2_W Q_f\right) \\
       g_{A1,f}&=&(x_b+s_W\gamma\sqrt{1-\gamma^2}x_a)\left(\frac{T_3}{2}\right)\\
      g_{A2,f}&=&(-x_a+s_W\gamma\sqrt{1-\gamma^2}x_b)\left(\frac{T_3}{2}\right)\\
      g_{\psi,1}&=& \sqrt{1-\gamma^2}x_a \\
       g_{\psi,2}&=& \sqrt{1-\gamma^2}x_b 
\end{eqnarray}
with $Q_f$ the SM fermion EM charge and $T_3$ its weak isospin third component. The resulting cross section is strongly suppressed by the mediator masses and by the small kinetic mixing parameter as expected. This suppression directly translates into a small annihilation rate both today and at the time of thermal freeze-out, if the symmetry breaking pattern is assumed to be identical to that at zero temperature.

 As we will discuss in the next section, the combination of $g_D^2 \epsilon^2/m_{A^{\prime}}^4$  is constrained by direct detection experiments. For example, for $m_{\psi} = 100$~GeV, the current bound implies $g_D^2\epsilon^2/m_{A^\prime}^4 \lesssim 10^{-19} $ (GeV)$^{-4}$. Such a small annihilation rate leads to an inefficient depletion of DM in the early Universe and consequently to a relic abundance far exceeding the observed value. For the parameters choice above, the predicted DM abundance is approximately seven orders of magnitude larger than the measured one.
This overabundance motivates considering a modified thermal history, in which the dark Higgs phase transition plays a central role in determining the final DM relic density.

We now consider the impact of a phase transition in the early Universe on the DM relic abundance. In order to keep the discussion as model independent as possible, we do not specify the microscopic dynamics responsible for the phase transition and instead focus on its macroscopic effects on the thermal history. In particular, although the dark gauge coupling $g_D$ generally affects the phase transition dynamics, its impact depends on the interplay with other parameters of the underlying model, such as the dark Higgs quartic and Yukawa couplings. We therefore do not assume a specific relation between $g_D$ and the phase transition parameters, treating the latter as free parameters. We further assume that the dynamics driving the phase transition does not directly affect the DM number density other than through the phase transition. Before the phase transition, the Universe is in the unbroken~(symmetric) phase of the dark $U(1)_D$ gauge symmetry, in which the dark photon is massless. DM annihilation takes place at temperatures for which the gauge symmetry remains unbroken. In this symmetric phase, kinetic mixing between the massless dark photon and the SM hypercharge gauge boson induces an effective mixing angle, allowing the dark sector, including dark fermions $\psi$ and the scalar field responsible for symmetry breaking, to remain in thermal equilibrium with the SM plasma.

We compute the thermally averaged annihilation cross section in the non relativistic limit, which is appropriate for thermal freeze-out. Since the dark Higgs phase transition has not yet occurred, the $U(1)_D$ gauge symmetry remains unbroken at these temperatures and the dark photon is massless. Once the temperature drops below the DM mass $m_{\psi}$, annihilation is dominated by the t- and u- channel processes $\psi\bar{\psi}\to A'A'$, with a thermally averaged cross section given by
\begin{equation}
\langle \sigma v\rangle= \frac{g_D^4   }{16\pi m^2_{\psi}}+\dots ,
\end{equation}
where $\dots$ represent subdominant terms from other annihilation channels. The dark photon remains in thermal equilibrium with the SM plasma through kinetic mixing interactions. DM thermal freeze-out occurs at a temperature $T_F\approx m_{\psi}/30$, and the resulting relic abundance is approximately
\begin{equation}
    \Omega_D h^2=\frac{2.5\times 10^{-10}\; {\rm GeV}^2}{\langle \sigma v \rangle}\, ,
    \label{eq:relic abudance}
\end{equation}
which is determined by the dark gauge coupling $g_D$ and the DM mass $m_{\psi}$. After DM has fallen out of thermal equilibrium and its abundance has frozen out, the $U(1)_D$ gauge symmetry is spontaneously broken through a first order phase transition. As a consequence, the dark photon acquires a non-zero mass, $m_{A^{\prime}} > m_\psi$, and the t- and u- channel annihilation processes into dark photons become kinematically forbidden. After the phase transition, the DM annihilation cross section is strongly suppressed, which would suggest an overabundant relic if one extrapolated from late time annihilation rates probed by indirect detection experiments. However, the relic abundance is already determined by the DM evolution in the symmetric phase. Moreover, such dark Higgs phase transitions are typically supercooled in the region of parameter space relevant for our scenario, as we will demonstrate explicitly in the next section. In this case, the vacuum energy released during the transition dominates over the radiation energy density, which can be quantified by the $\alpha$ parameter, $\alpha = \rho_{vac}/\rho_{rad}$. In the supercooled region, $\alpha \gg 1$, so that the completion of the phase transition reheats the Universe and injects a substantial amount of entropy. The entropy injection associated with the supercooled phase transition dilutes the DM abundance~\cite{Hambye:2018qjv,Roy:2022gop}. The dilution factor is given by the ratio of comoving entropies before and after reheating, $(T_{rh}/T_{p})^3$, where $T_p$ is the percolation temperature at which true vacuum bubbles become infinitely connected, and $T_{rh}$ is the reheating temperature. Using energy conservation during reheating $T_{rh} = T_p (1+\alpha)^{1/4}$, with $\alpha$ he usual strength parameter of the phase transition, the resulting dilution factor becomes
\begin{equation}
    \Delta = \left(\frac{T_{rh}}{T_p}\right)^3 = (1+\alpha)^{3/4}.
\end{equation}
The final relic density is thus given by the freeze-out value divided by $\Delta$. The overabundance that can be accommodated is bounded by two considerations, the lower thermalization bound on $\epsilon$, which limits how large the initial overabundance can be, and the entropy dilution factor $(1+\alpha)^{3/4}$, which determines how much of this abundance can be reduced during the phase transition.

\section{Phenomenological Constraints and Signals}
\label{sec:constraints and signals}

We now discuss the experimental and cosmological constraints relevant for the model. One constraint to the model comes from direct detection searches of DM. We calculate the spin-independent DM–proton and DM–neutron scattering cross sections, which proceed via t-channel exchange of the neutral mediators $\chi_{1,2}$. The corresponding cross sections are given by,
\begin{eqnarray}
    \sigma_{\psi p}&=&\frac{g^2_D g^2 \mu^2_{\psi p}}{\pi c^2_W m^4_{1}m^4_{2}}\left((2 g_{V1,u}+g_{V1,d})g_{\psi,1}m^2_{2}+(2 g_{V2,u}+g_{V2,d})g_{\psi,2}m^2_{1}  \right)^2 \\
    \sigma_{\psi n}&=&\frac{g^2_D g^2 \mu^2_{\psi n}}{\pi c^2_W m^4_{1}m^4_{2}}\left(( g_{V1,u}+2g_{V1,d})g_{\psi,1}m^2_{2}+( g_{V2,u}+2g_{V2,d})g_{\psi,2}m^2_{1}  \right)^2
\end{eqnarray}
where $\mu_{\psi i}=m_{\psi}m_{i}/(m_{\psi}+m_i)$, with $i=p,n$ denotes the reduced mass of the of the DM–nucleon system. Note that as expected, only the vectorial couplings to SM matter fields appear. From these expressions we obtain the spin independent DM-nucleon scattering cross section,
\begin{equation}
    \sigma_{\psi N}=\frac{Z\sigma_{\psi p}+(A-Z)\sigma_{\psi n}}{A}
\end{equation}
where $Z$ is the atomic number and $A$ the atomic mass of the medium in which the DM scatters off.

Another constraint comes from precision electroweak measurements, which imply $\epsilon \lesssim 10^{-2}$ for a light dark photon mass $m_{A^{\prime}}$ sufficiently away from the $Z$-mass, with the bound becoming weaker for $m_{A^{\prime}}\gg m_Z$ ~\cite{Langacker:2008yv,Curtin:2014cca}. These bounds may be relaxed in the presence of sizable invisible decay widths, though they nonetheless restrict the viable parameter space.

Consistency of the thermal history requires that the dark sector was in thermal equilibrium with the SM plasma prior to DM freeze-out. Requiring the interaction rate for processes such as $\psi \psi \rightarrow \textrm{SM}\textrm{SM} $ to exceed the Hubble expansion rate leads to a lower bound on the kinetic mixing. Roughly speaking, the interaction rate scales as $\sim \epsilon^2\alpha_{EM}\alpha_Dm\psi$, where $\alpha_{EM}=e^2/4\pi$, which translates into the approximate condition
\begin{equation}
    \epsilon g_D \gtrsim 10^{-7} \sqrt{\frac{m_{\psi}}{\textrm{GeV}}}.
    \label{eq:equilibrium}
\end{equation}
In addition, kinetic mixing in the presence of a non-vanishing $m_{A'}$ and of EWSB, leads to the mixing described in the previous paragraphs for the neutral gauge bosons of the dark and EW sectors. The dark Higgs decays, via off-shell gauge bosons, to SM fermions as: $\varphi\to \chi^{*}\chi^{*}\to f\bar{f}f'\bar{f'}$, with $\chi=\chi_{2\mu}, \chi_{1\mu}$, and we must demand that these decays happen at times earlier than Big Bang nucleosynthesis~(BBN), in order not to disrupt the well-measured primordial abundances of the early Universe. In addition, successful percolation imposes an additional requirement on strongly supercooled transitions. While larger supercooling can increase the released vacuum energy and hence the entropy dilution, it also delays and may prevent the transition from completing. Therefore, regions with stronger supercooling, corresponding to larger $\alpha$, the completion of the phase transition must be checked. In our model-independent treatment we assume that the benchmark phase transitions complete successfully. A quantitative percolation analysis would require specifying the underlying scalar potential and computing the corresponding tunneling rate.
We now introduce some representative benchmark scenarios for this framework, summarized in Table \ref{tab:BMs}. We first consider DM mass in the range $m_{\psi}$ = 10, 30, 100, and 1000~GeV, spanning the region where direct detection constraints vary significantly and are particularly stringent around $m_{\psi}\simeq 30$~GeV. For each benchmark point, we first determine the dark gauge coupling $g_D$ by requiring the obtained relic abundance using Eq.~(\ref{eq:relic abudance}) matches the observed value of  $\Omega h^2 = 0.12$~\cite{Planck:2018vyg}. When computing the relic abundance, we fix the $\alpha$ parameter, which controls the amount of entropy dilution associated with the dark Higgs phase transition to $\alpha = 10$. Larger values of $\alpha$ lead to stronger dilution and therefore require smaller values of $g_D$. The relic abundance is diluted by a factor of $(1+\alpha)^{3/4} \simeq \alpha^{3/4}$ for $\alpha \gg1$. As a result, to get the observed relic density implies that the dark gauge coupling rescales as as $g_D \rightarrow g_D \alpha^{3/16}$. For instance, taking $\alpha = 10$ as the reference value, increasing $\alpha$ to $10^2$, $10^3$, $10^4$ reduces the required value of the dark gauge coupling by a factor of approximately 1.5, 2.4, and 3.6 respectively. The kinetic mixing $\epsilon$ is chosen to be the minimum value required to keep the dark sector in thermal equilibrium with the SM in the early Universe. Using Eqs.~(\ref{eq:equilibrium}) and (\ref{eq:relic abudance}), this yields $\epsilon \simeq 9\times10^{-6}$ for $\alpha = 10$. For larger supercooling, the reduced value of $g_D$ increases the minimum kinetic mixing $\epsilon$ required for thermalization by the same factor. As discussed in Sec~\ref{sec:relic abundance}, achieving the observed relic density in this scenario requires the dark Higgs phase transition to occur after DM freeze-out. We therefore choose the phase transition temperature to be roughly $T_{*}\sim T_F/3\sim m_{\psi}/100$. This serves as a representative benchmark rather than a special value. The requirement is simply that the phase transition occur after freeze-out, $T_* < T_F$, so that the relic abundance is initially determined in the symmetric phase. Lower values of $T_*$ correspond to stronger supercooling, leading to more entropy production at reheating and thus additional dilution of the DM abundance. Therefore, the dark gauge coupling $g_D$ required to reproduce the observed relic density can be further reduced. One special region of parameter space is when the freeze-out temperature and the phase transition temperature becomes comparable. Nevertheless, if this situation is realized, the relic abundance can be affected in several ways. In particular, the effective annihilation rate during freeze-out, averaged over the coexisting phases, would be smaller, leading to a larger final relic abundance. In addition, if the phase transition takes place at a temperature comparable to the freeze-out temperature, ($T_{\rm *}\sim T_{\rm F}$), the amount of supercooling is typically reduced. This would also imply a smaller entropy injection from the transition, and hence a weaker dilution of the DM abundance. These effects can quantitatively modify the final relic density, although a dedicated treatment of freeze-out during the bubble nucleation epoch lies beyond the scope of the present work. The dark photon mass $m_{A^{\prime}}$ is taken to be slightly above the DM mass $m_{\psi}$. The hierarchy of $m_{A^{\prime}} > m_{\psi} \gg T_*$ implies a large hierarchy between the dark photon mass and the phase transition temperature, indicating a strongly first order transition accompanied by significant supercooling.  The corresponding cross sections are listed in Table \ref{tab:BMs}, and they are within the current direct detection limit~\cite{XENON:2017vdw,LZ:2024zvo}. Direct detection is largely insensitive to the dilution from supercooling once the observed relic abundance is imposed. Direct detection cross section scales as $\epsilon^2 g_D^2/m_{A^{\prime}}^4$. Since $g_D$ and 
$\epsilon$ rescale in a complementary way, the direct detection rate remains nearly unchanged when varying $\alpha$. Finally, for each benchmark we compute the DM annihilation cross section today to illustrate that, in the absence of the late-time phase transition, the model would predict a severely overabundant DM relic density. Similar to direct detection, the annihilation cross section today is also insensitive to additional supercooling. It scales as $g_D^2\epsilon^2/m_{A^{\prime}}^2$, and therefore remains unchanged under the complementary rescaling of $g_D$ and $\epsilon$. 


\begin{table}
    \centering
    \begin{tabular}{|c|cccc|}
        \hline
        & BMA & BMB  & BMC & BMD\\
        \hline
        $m_\psi$ (GeV) & 10 & 30 & 100 & 1000\\
         $g_D$& 0.036 & 0.063& 0.11 & 0.36\\
         $m_{A^{\prime}}$ (GeV)& 18 &50 &110 &1100 \\
         \hline
        $\Omega h^2$&0.12 &0.12 &0.12 &0.12\\
        $\sigma_{SI}$ (cm$^2$) & $8.3\times 10^{-48}$ &$5.1\times 10^{-49}$ & $1.1\times 10^{-49}$&$1.1 \times 10^{-53}$  \\
         $\langle \sigma v\rangle_{\textrm{today}}$ (cm$^3$ s$^{-1}$) & 3.0$\times 10^{-34}$& 5.6$\times 10^{-35}$ & $2.0\times 10^{-36}$  &$2.0\times 10^{-37}$\\
        \hline
    \end{tabular}
    \caption{Benchmark scenarios considered in this work. We assume $\alpha = 10$ for all benchmarks. For each benchmark, we list the DM mass $m_\psi$, the dark gauge coupling $g_D$ fixed by the relic abundance requirement, and the dark photon mass $m_{A^{\prime}}$. The kinetic mixing $\epsilon$ is fixed to be $9\times 10^{-6}$. The resulting DM relic abundance $\Omega h^2$ and spin-independent DM–nucleon scattering cross section $\sigma_{SI}$, and the DM annihilation cross section today $\langle \sigma v\rangle_{\textrm{today}}$ are also shown. All benchmarks satisfy current direct detection constraints.}
    \label{tab:BMs}
\end{table}

We now use the benchmark scenarios introduced above to discuss the gravitational wave~(GW) signal from the dark Higgs phase transition. A first order phase transition releases latent heat into the plasma and generically produces a stochastic GW background~\cite{Jinno:2016knw,Caprini:2015zlo,Caprini:2019egz,Hindmarsh:2016lnk,Hindmarsh:2019phv}. The dominant sources of GWs in such a transition are sound waves in the plasma, collisions of expanding true-vacuum bubbles, and magnetohydrodynamic turbulence. In most scenarios, the contribution from turbulence is subdominant, while the contribution from bubble collisions is subject to significant theoretical uncertainties~\cite{Caprini:2019egz,Hindmarsh:2015qta,Lewicki:2020jiv}. In this work, we therefore focus on the GW signal sourced by sound waves in the plasma, and include the contribution from turbulence for completeness. The GW spectrum generated by a first order phase transition is characterized by four macroscopic parameters: the strength of the phase transition $\alpha$, which also controls the dilution due to entropy injection, the mean bubble separation normalized by the Hubble radius $R_*H_*$, the bubble wall velocity $v_w$, and the phase transition temperature $T_*$. In the benchmark scenarios considered here, the dark Higgs vev is much larger than the phase transition temperature, as can be inferred from the dark photon mass $m_{A^{\prime}}$. This hierarchy implies a large ratio of $v_{\Phi}/T_*$ and therefore a supercooled phase transition.

\begin{figure}[t]
  \centering
  \begin{subfigure}{0.38\textwidth}
    \includegraphics[width=\linewidth]{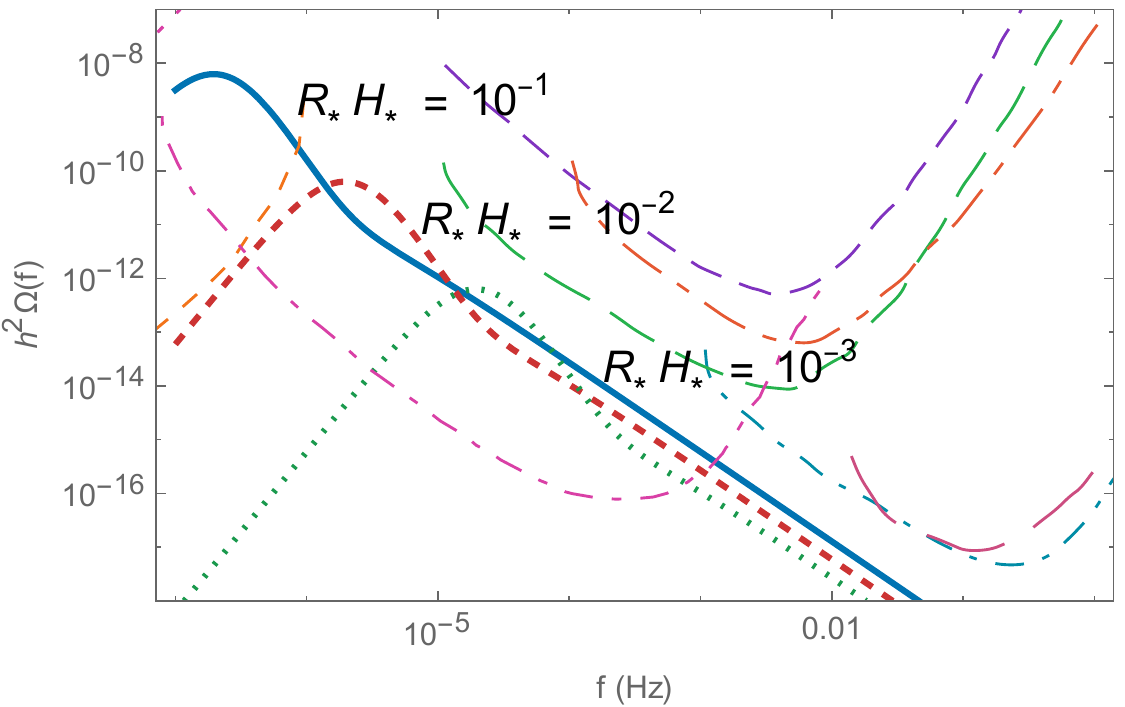}
    \caption{$m_\psi$ = 10 GeV}
  \end{subfigure}
  \begin{subfigure}{0.38\textwidth}
    \includegraphics[width=\linewidth]{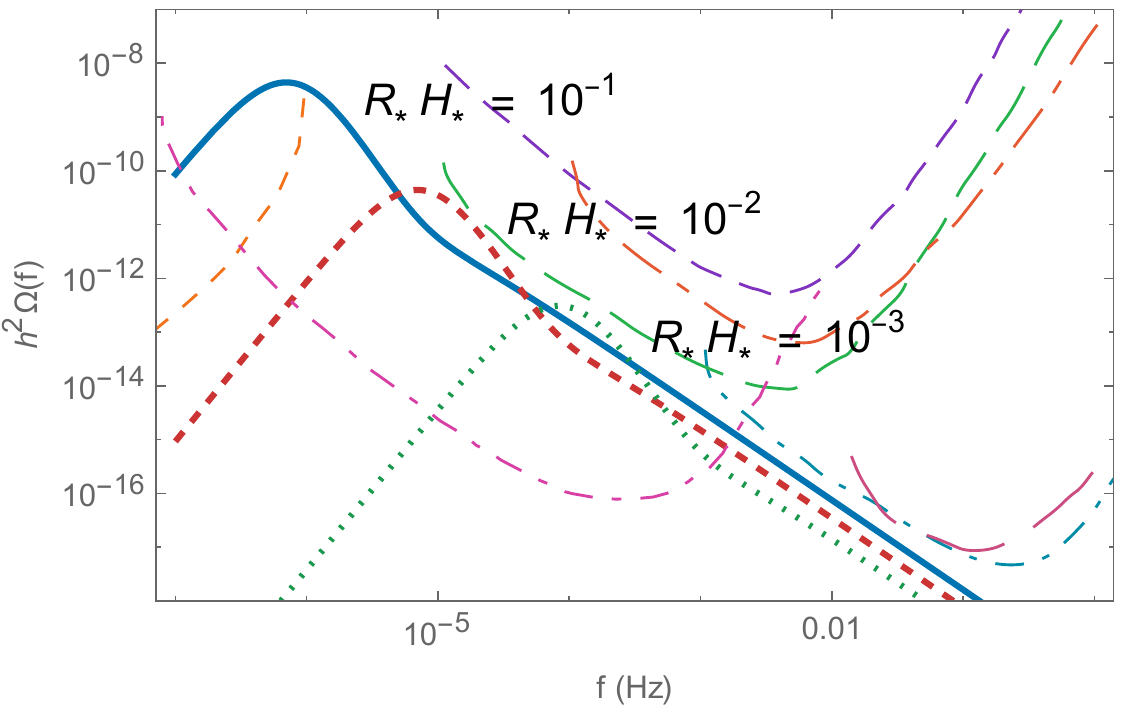}
    \caption{$m_\psi$= 30 GeV}
  \end{subfigure}

  \medskip

  \begin{subfigure}{0.38\textwidth}
    \includegraphics[width=\linewidth]{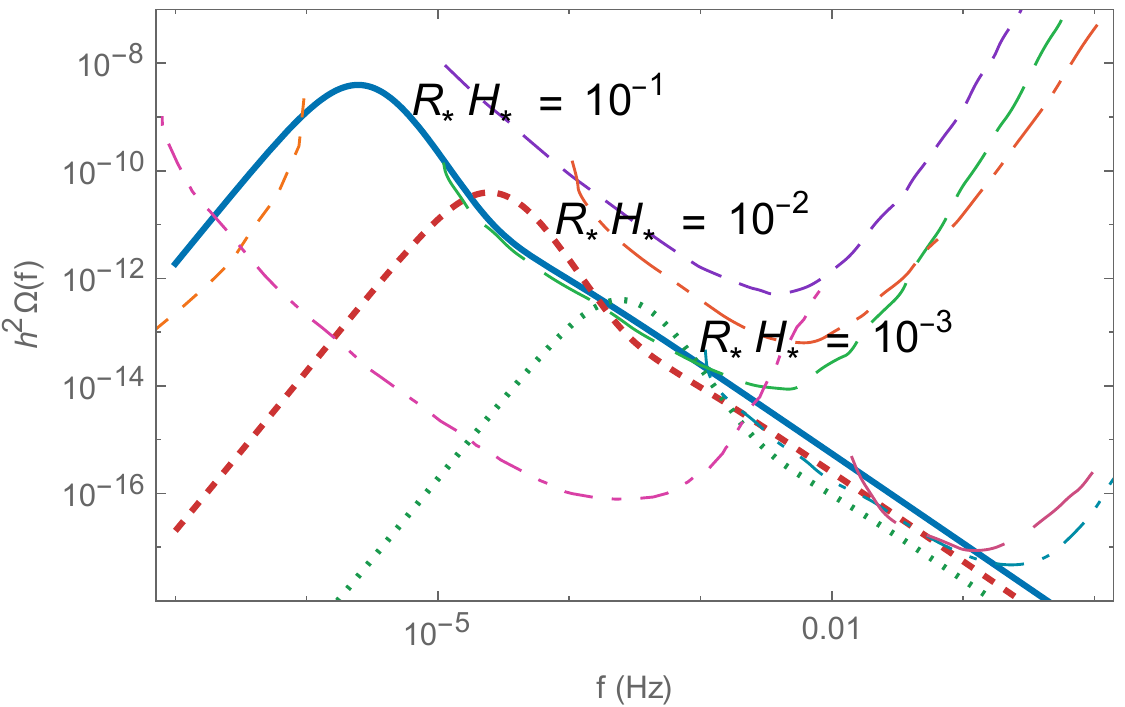}
    \caption{$m_\psi$ = 100 GeV}
  \end{subfigure}
  \begin{subfigure}{0.38\textwidth}
    \includegraphics[width=\linewidth]{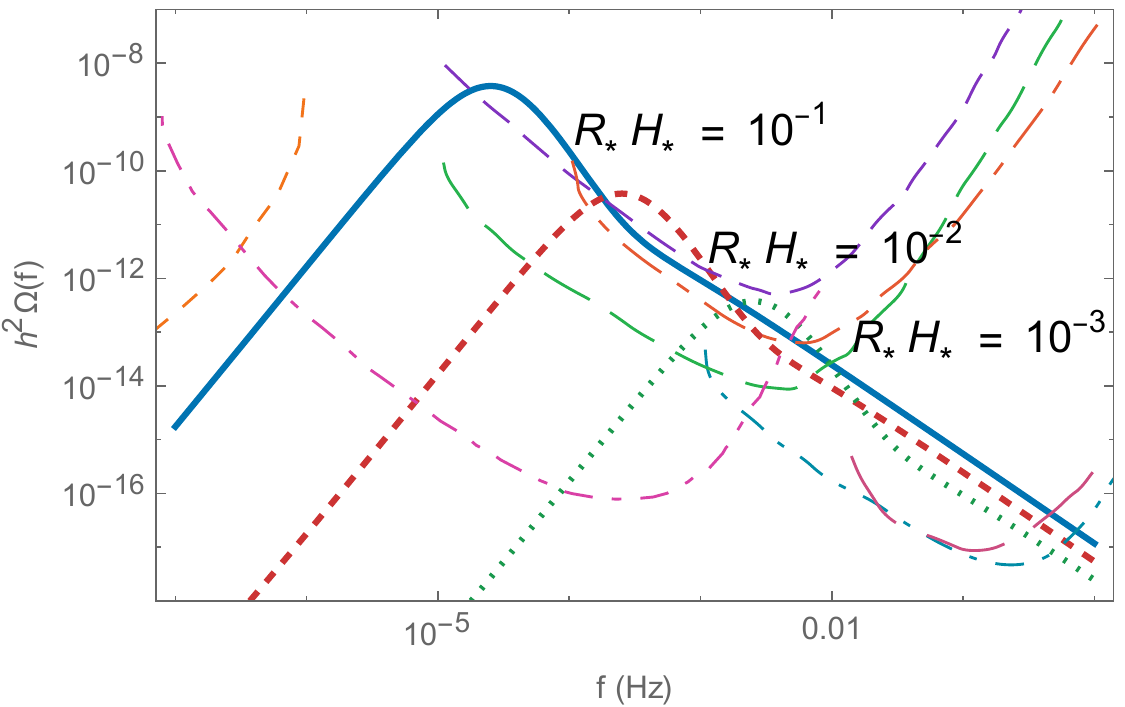}
    \caption{$m_\psi = 1000$ GeV}
  \end{subfigure}
  \hspace{0.02\textwidth}
  \begin{subfigure}{0.18\textwidth}
    \includegraphics[width=\linewidth]{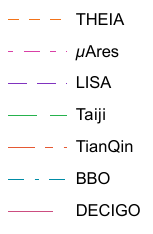}
    \caption*{}
  \end{subfigure}

  \caption{GW spectra generated by the dark Higgs first order phase transition for the benchmark scenarios BMA–BMD listed in Table~\ref{tab:BMs}. Each panel corresponds to a different DM mass and model parameters as specified in the table. The GW signal is computed assuming a supercooled phase transition with phase transition temperature $T_*\simeq m_\psi/100$, $\alpha$ parameter $\alpha = 10$, bubble wall velocity $v_w=1$. The blue solid, red dashed, and green dotted curves correspond to different values of the mean bubble separation $R_*H_* = 10^{-1}$, $10^{-2}$, and $10^{-3}$ respectively. Projected sensitivity curves for future GW observatories, including THEIA~\cite{2018FrASS...5...11V}, $\mu \rm{Ares}$~\cite{Sesana:2019vho}, LISA~\cite{Caprini:2019egz,amaro2017laser,Robson:2018ifk}, Taiji~\cite{Ruan:2018tsw}, TianQin~\cite{TianQin:2015yph}, DECIGO~\cite{Kawamura:2020pcg,Kawamura:2006up}, and BBO~\cite{Corbin:2005ny,Yagi:2011wg} are shown for comparison}
  \label{fig:benchmarks}
\end{figure}

Supercooled phase transitions are typically associated with large values of the strength parameter $\alpha$. Since the GW amplitude scales approximately as $(\alpha/(\alpha+1))^2$ the signal saturates for $\alpha \gg 1$, allowing us to neglect the dependence on $\alpha$ in this regime. As emphasized in Ref.~\cite{Athron:2022mmm,Ellis:2018mja,Ellis:2019oqb,Ellis:2020nnr}, for strongly supercooled phase transitions the commonly used parameter $\beta/H_*$, which characterizes the inverse duration of the transition, is no longer an optimal descriptor of the dynamics. Instead, the mean bubble separation $R_*H_*$ provides a more appropriate characterization of the GW signal. We therefore adopt $R_* H_*$ as a key parameter in our analysis and treat it as a free parameter, which can be computed once the underlying particle physics model of the phase transition is fully specified. The phase transition temperature is taken to be $m_\psi/100$ as discussed before. Given that the peak frequency today scales linearly with $T_*$, decreasing the phase transition temperature shifts the signal toward lower frequencies. In the case of supercooled phase transitions, this temperature corresponds to the reheating temperature, which is typically identified as the effective time of GW production. Moreover, bubbles in supercooled transitions are expected to accelerate to ultra-relativistic velocities. We therefore take the bubble wall velocity to be $v_w = 1$. Following the formalism of Ref.~\cite{Lewicki:2020jiv}, we compute the resulting GW spectra for our benchmark scenarios. The corresponding GW signals are shown in Fig.~\ref{fig:benchmarks}, where we also overlay the projected sensitivities of upcoming GW experiments. We fix $\alpha = 10$ in the calculation. The signal increases moderately with stronger supercooling and is enhanced by approximately $20\%$ in the limit $\alpha \to \infty$. The values of $R_{*}H_{*}$ and $\alpha$ used for the figures are those expected for a supercooled phase transition~\cite{Ellis:2018mja,Ellis:2019oqb}. Note that, for fixed $m_{\psi}$, as one decreases the bubble separation $R_{*}H_{*}$, the GW spectra peak decreases and shifts towards larger frequency values. On the other hand, for fixed $R_{*}H_{*}$, larger values of the DM mass $m_{\psi}$ shift all of the GW spectra to larger frequencies. The latter is a consequence of how we are relating the phase transition temperature to the DM mass. As we see from Fig.~\ref{fig:benchmarks}, many of the upcoming and planned GW experiments should be able to probe these DM scenarios for different DM masses.

We now briefly comment on the scenario in which the DM is light, with mass 
$m_\psi \sim O(1)$ GeV. This regime is particularly well motivated, as light DM has attracted considerable attention in recent years. One of the strongest constraints on DM with masses below or around the GeV scale comes from late time annihilation into electromagnetically charged particles, which injects energy into the primordial plasma after recombination and modifies the ionization history, leaving observable signatures in the CMB. CMB observations typically constrain the velocity-averaged annihilation cross section to be $\langle \sigma v\rangle \lesssim 10^{-28} -10^{-29} \textrm{cm}^3 \textrm{s}^{-1}$ for DM masses of order a GeV, which is several orders of magnitude smaller than the canonical thermal relic value $\langle \sigma v\rangle \simeq 3\times 10^{-26} \textrm{cm}^3 \textrm{s}^{-1}$. Several approaches have been explored in the literature to accommodate light DM while remaining consistent with these constraints. In this work, we show that a late first order phase transition in the dark sector provides an additional way to suppress late-time DM annihilation.

DM in the GeV range is also interesting from the perspective of GW phenomenology. A dark sector phase transition occurring at temperatures in the range $T_* \sim 10 - 100$ MeV, corresponding to DM masses $m_\psi \sim 1 - 10 $ GeV, produces a GW signal peaking in the nano-Hertz frequency band. This signal lies in the sensitivity range of the pulsar timing array experiments and has been discussed as a possible explanation of the NANOGrav observations \cite{NANOGrav:2020bcs,NANOGrav:2023gor}.

\begin{figure}
    \centering
    \includegraphics[width=0.70\linewidth]{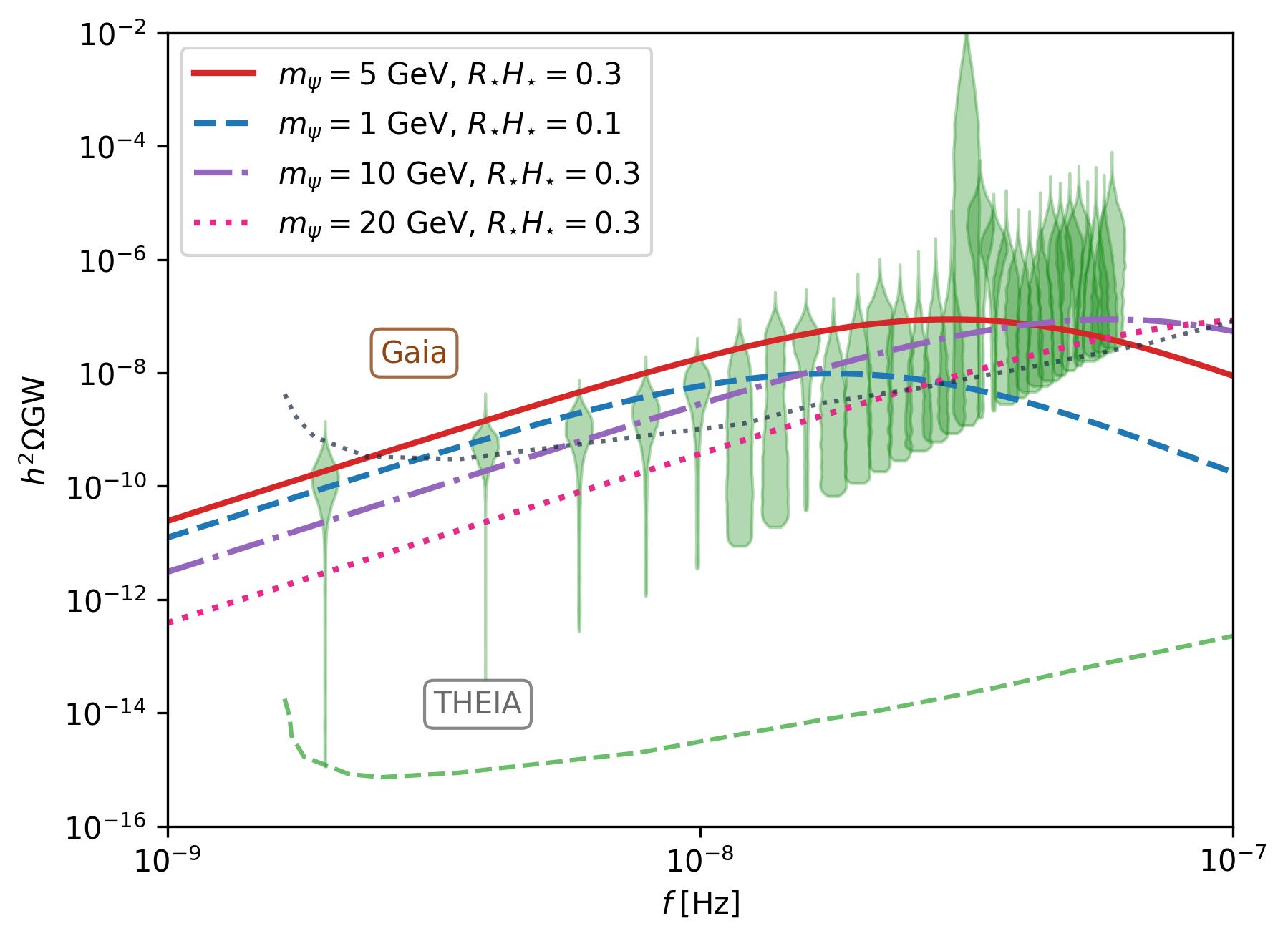}
    \caption{GW spectra from a supercooled dark Higgs phase transition for different DM masses and phase transition temperatures, overlaid with the NANOGrav 15-year data set~\cite{NANOGrav:2023gor} and the projected sensitivity of Gaia~\cite{brown2018gaia}, and THEIA~\cite{2018FrASS...5...11V}. All spectra are computed assuming strong supercooling with $\alpha = 100$, relativistic bubble wall, $v_w\simeq 1$  and values of the mean bubble separation parameter $R_{*} H_{*}$ in the range $0.1$–$0.3$, as indicated. The phase transition temperature is assumed to be $T_{*} \simeq m_{\psi}/100$. }
    \label{fig:DM5}
\end{figure}

As a concrete example, we consider four benchmark scenarios with DM mass ranges between 1 - 20~GeV. The resulting GW spectrum is shown in Fig.~\ref{fig:DM5}, overlaid with the NANOGrav 15 year data~\cite{NANOGrav:2023gor}, and the projected sensitivity from Gaia~\cite{brown2018gaia} and THEIA~\cite{2018FrASS...5...11V}. 
All curves are computed assuming strong supercooling with $\alpha = 100$, which enhances the signal and improves the fit to the NANOGrav data, and span values of $R_* H_*$ in the range $0.1$-$0.3$, as labeled. The $m_\psi = 5\,\mathrm{GeV}$ case (red solid curve), corresponding to $T_* = 50\,\mathrm{MeV}$ and $R_* H_* = 0.3$, provides a viable fit to the NANOGrav signal. The dot-dashed magenta curve corresponds to a heavier DM mass, $m_\psi = 10\,\mathrm{GeV}$, or equivalently a higher phase transition temperature $T_* = 100\,\mathrm{MeV}$. In this case, the peak frequency shifts to higher values and the overall signal amplitude is reduced. We therefore choose $R_* H_* = 0.3$ to enhance the signal and lower the peak frequency so that it also provides a viable fit to the NANOGrav signal. For the $m_\psi=5\, \mathrm{GeV}$ scenario, the peak frequency of the GW spectrum lies within the observational band of NANOGrav, which results in a signal that is slightly on the higher end of the NANOGrav data in the low frequency bins, and slightly on the lower end of the highest frequency bins. Then for $m_\psi = 10\,\mathrm{GeV}$ case, the peak frequency shifts toward higher frequencies, leading to a signal that lies somewhat on the lower end of the in the lowest-frequency bins. For lighter DM with $m_\psi = 1\,\mathrm{GeV}$ ($T_* = 10\,\mathrm{MeV}$), the peak frequency shifts to lower values and the signal amplitude increases, allowing a fit near the peak with a smaller bubble separation parameter, $R_* H_* = 0.1$. However, the spectrum falls off too steeply at higher frequencies and fails to reproduce the signal in the higher frequncy bins. For heavier DM, $m_\psi = 20\,\mathrm{GeV}$ ($T_* = 200\,\mathrm{MeV}$), the peak frequency moves well above the NANOGrav sensitivity range and the signal becomes too weak in the low frequency bins, even for $R_* H_* = 0.3$. These results indicate that phase transition temperatures in the range $T_* \sim 50$–$100\, \mathrm{MeV}$, corresponding to DM masses $m_\psi \sim 5$–$10\,\mathrm{GeV}$ in our setup, provide possible fits to the NANOGrav signal. 

For DM in the few-GeV mass range interacting through a light kinetically mixed dark photon, existing direct detection constraints are comparatively weak. Nuclear recoil experiments rapidly lose sensitivity due to detector energy thresholds, while electron recoil searches, collider experiments, fixed-target experiments, and beam-dump experiments are only beginning to probe this region of parameter space. As a result, models featuring light dark photons with small kinetic mixing remain weakly constrained at present, but will be tested by a variety of upcoming experiments in the near future. In this context, the GW signal from a dark sector phase transition provides a complementary and potentially powerful probe of light DM.


\section{Conclusion}
\label{sec:conclusion}
In this work, we studied a dark sector consisting of a kinetically mixed dark photon, a dark Higgs field undergoing a first-order phase transition, and fermionic Dirac DM whose mass remains unchanged during the transition. We showed that although the DM annihilation rate is strongly suppressed at late times, leading to a naive expectation of overabundance, the relic density is instead set in the symmetric phase before the dark Higgs phase transition. Before the transition, the dark photon is massless and DM annihilation is dominated by the unsuppressed t- and u-channel processes $\psi\bar{\psi}\rightarrow A^{\prime} A^{\prime}$, which efficiently deplete the DM abundance. After the phase transition, the dark photon acquires a mass, these channels become kinematically forbidden, and annihilation occurs only through s-channel dark photon exchange into SM, which is suppressed by the small kinetic mixing. As a result, the correct relic abundance is obtained while the DM annihilation cross section at late times remains strongly suppressed, which is particularly important for satisfying CMB constraints for light DM. We identified representative benchmark scenarios  consistent with direct detection bounds, electroweak precision tests, and CMB constraints. 

A key observable of this framework is a stochastic GW background generated by the dark Higgs phase transition, which is typically supercooled in the parameter space of interest. We showed the resulting signal for this scenario can be probed by future GW observatories. For light DM, the resulting GW signal can fall in the nano-Hertz frequency range and may account for the signal observed by NANOGrav.

In conclusion, our results show how early-Universe phase transitions can play an important role in determining DM properties and demonstrate that GW signals offer a complementary way to probing dark sectors. 

\section*{Acknowledgements}
We thank S. Gori and S. Roy for helpful discussions. The work of CW at the University of Chicago has been supported by the DOE grant DE-SC0013642. C.W. would like to thank the Aspen Center for Physics, which is supported by National Science Foundation grant No. PHY-1607611, where part of this work has been done.
The research of A.M. was supported in part by Perimeter Institute for Theoretical Physics.
Research at Perimeter Institute is supported by the Government of Canada through the
Department of Innovation, Science, and Economic Development, and by the Province
of Ontario through the Ministry of Colleges and Universities. The work of PH is supported by the National Science
Foundation under grant number PHY-2412875, and the University Research Association Visiting Scholars Program. PH is grateful for the hospitality of Perimeter Institute where part of this work was carried out. Research at Perimeter Institute is supported in part by the Government of Canada through the Department of Innovation, Science and Economic Development and by the Province of Ontario through the Ministry of Colleges and Universities. This work was supported by a grant from the Simons Foundation (1034867, Dittrich). 

\bibliographystyle{biblio/bibstyle}
\bibliography{biblio/biblio.bib}

\end{document}